\documentclass{jfm}
\usepackage{eurosym}
\usepackage{amsmath}
\usepackage{graphicx}
\usepackage{natbib}
\usepackage{upmath}
\usepackage{amssymb}
\usepackage{amsbsy}
\usepackage{pdfpages}
\usepackage{subfigure}
\usepackage{float}
\usepackage{caption}
\usepackage{xcolor}
\usepackage{caption}
\usepackage[T1]{fontenc}

\DeclareMathAlphabet\mathbfcal{OMS}{cmsy}{b}{n}
\ifCUPmtlplainloaded \else
  \checkfont{eurm10}
  \iffontfound
    \IfFileExists{upmath.sty}
      {\typeout{^^JFound AMS Euler Roman fonts on the system,
                   using the 'upmath' package.^^J}
}
      {\typeout{^^JFound AMS Euler Roman fonts on the system, but you
                   dont seem to have the}
       \typeout{'upmath' package installed. JFM.cls can take advantage
                 of these fonts,^^Jif you use 'upmath' package.^^J}
       
      }
  \else
    
  \fi
\fi
\ifCUPmtlplainloaded \else
  \checkfont{msam10}
  \iffontfound
    \IfFileExists{amssymb.sty}
      {\typeout{^^JFound AMS Symbol fonts on the system, using the
                'amssymb' package.^^J}

      }{}
  \fi
\fi
\ifCUPmtlplainloaded \else
  \IfFileExists{amsbsy.sty}
    {\typeout{^^JFound the 'amsbsy' package on the system, using it.^^J}
}
    {\providecommand\boldsymbol[1]{\mbox{\boldmath $##1$}}}
\fi

\newsavebox{\astrutbox}
\sbox{\astrutbox}{\rule[-5pt]{0pt}{20pt}}

\DeclareMathAlphabet{\mathbfsf}{\encodingdefault}{\sfdefault}{bx}{sl}

\DeclareMathAlphabet{\mathsfit}{\encodingdefault}{\sfdefault}{m}{sl}

\newcommand{\tens}[1]{\mathbfsf{#1}}
\newcommand{\te}[1]{\mathsfit{#1}}

\affiliation{$^1$School of Civil and Environmental Engineering, Cornell University,
Ithaca, NY 14853, USA\\[\affilskip]
$^2$Wenzhou Institute, University of Chinese Academy of Sciences, Wenzhou, Zhejiang 325001, China
\\[\affilskip]
$^3$Dipartimento di Scienze dell'Ingegneria Civile e dell'Architettura, Politecnico di Bari, 70125 Bari, Italy}
\pubyear{2010}
\volume{650}
\pagerange{119--126}

\begin{document}

\title[Planar extensional flows of a dense suspension]{ Predictions of microstructure and
stress in planar extensional flows of a dense viscous suspension}
\author[ ]{James T. Jenkins$^1$ \thanks{%
Email address for correspondence regarding the model, jtj2@cornell.edu}, Ryohei Seto$^2$ \thanks {Email address for correspondence regarding the simulation, seto@wiucas.ac.cn} and Luigi La Ragione$^3$}
\date{?; revised ?; accepted ?. - To be entered by editorial office}
\maketitle

\begin{abstract}
We consider extensional flows of a dense layer of spheres in a viscous fluid and employ force and torque balances to determine the trajectory of particle pairs that contribute to the stress. In doing this, we use Stokesian dynamics simulations to guide the choice of the near-contacting pairs that follow such a trajectory. 
We specify the boundary conditions on the representative trajectory, and determine the distribution of particles along it and how the stress depends on the microstructure and strain rate. We test the resulting predictions using the numerical simulations. Also, we show that the relation between the tensors of stress and strain rate involves the second and fourth moments of the particle distribution function. 
\end{abstract}

\begin{keywords}
Authors should not enter keywords on the manuscript, as these must be chosen by the author during the online submission process and will then be added during the typesetting process (see http://journals.cambridge.org/data/\linebreak[3]relatedlink/jfm-\linebreak[3]keywords.pdf for the full list)
\end{keywords}

\section{Introduction}
In a recent study, Jenkins and La Ragione (2015) determine the typical trajectory of a  force-equilibrated pair of particles of a dense, two-dimensional suspensions of spheres subjected to a simple shearing flow. They evaluate the distribution function of such near-contacting neighbours along the trajectory and, using this distribution function and the expression for the force between the pair, they predict the particle pressure, the difference in normal stresses and the difference between the average rotation of the spheres and half the vorticity of the average velocity.

Here, we focus on extensional flows, also called pure shearing, of a dense layer of spheres and, as an extension of the previous work, also introduce the moment equilibrium. We employ a simplified Stokesian dynamics numerical simulation, perhaps more properly called lubrication dynamics (e.g., Ball and Melrose 1995), to guide the choice of the near-contacting pairs on a representative trajectory that contributes most to the inter-particle stress. We specify the boundary conditions on the representative trajectory, and determine the distribution of particles along it and the relationship between stress, microstructure and strain rate. We test these predictions against the results of the numerical simulations. We show that the relation between the stress and strain rate tensors involves the second and fourth moments of the particle distribution, and place this and other aspects of our approach in the context of the earlier models of Phan-Thien (1995), Stickel, et al. (2006), Goddard (2006), Gillissen and Wilson (2018, 2019) and Gilissen et al. (2019) that focus on the second moment and that of Chacko et al. (2018), who introduce a fourth-rank tensor to describe flow reversal.

The approximate satisfaction of force and torque balances for particles in the flow plays an important role in what we do. In that regard, we operate in the spirit of Nazockdast and Morris (2012a, 2012b, 2013) or that of the statistical characterization by Thomas et. al. (2018) of equilibrated particles sheared in two dimensions, but in the limit of dense flows of the planar extensional flow. The analysis must be extended to three-dimensional simple shear flows before it can be placed in relation to phenomenological relations that have resulted from experiments on dense three-dimensional shearing flows (Boyer, et al. 2011; Guazzelli and Pouliquen 2018).

\section{Micro-mechanics}

A steady, planar extensional flow of a dense suspension of identical spheres with radius $a$ is characterized
by an average rate of deformation tensor $\tens{D}$ with non-zero
components $\te{D}_{11}=-\te{D}_{22}=\dot{\gamma}$, where $x_{1}$ and $x_{2}$ are
the axes in the directions of greatest extension and compression,
respectively, and $\dot{\gamma}$ is the constant shear rate. We focus on a
typical pair of spheres and their near-contacting neighbours, and take 
$ \boldsymbol{\hat{d}}^{(BA)}$ 
to be the unit vector directed from the centre of sphere $A$ to that of sphere $B$, 
with $\boldsymbol{\hat{d}}^{(AB)} = -\boldsymbol{\hat{d}}^{(BA)}$ 
(see Fig. 1). Then, with $\theta^{(BA)}$ the time-dependent angle
between $\boldsymbol{\hat{d}}^{(BA)}$ and the $x_{2}$ axis, 
\begin{equation}
\hat{d}_{\alpha}^{(BA)}=(\sin \theta^{(BA)},\cos \theta^{(BA)})
\end{equation}%
and the components of the unit tangent vector, 
$\boldsymbol{\hat{t}}^{(BA)}=-\boldsymbol{\hat{t}}^{(AB)}$, perpendicular to it, are
\begin{equation}
  t_{\alpha }^{(BA) }=( \cos \theta^{(BA)},-\sin \theta^{(BA)}),
\end{equation}%
or $t_{\alpha }^{(BA)} = \varepsilon_{\alpha \beta} d_{\beta }^{(BA)}$, 
where $\varepsilon_{12} = -\varepsilon_{21} = 1$ and $ \varepsilon_{11} = \varepsilon_{22} =0$. The unit vectors $\boldsymbol{\hat{d}}^{(BA)}$ and $\boldsymbol{\hat{t}}^{(BA)}$ are indicated in Fig. 1.

\begin{figure}
  \centering
  \includegraphics[scale=0.50]{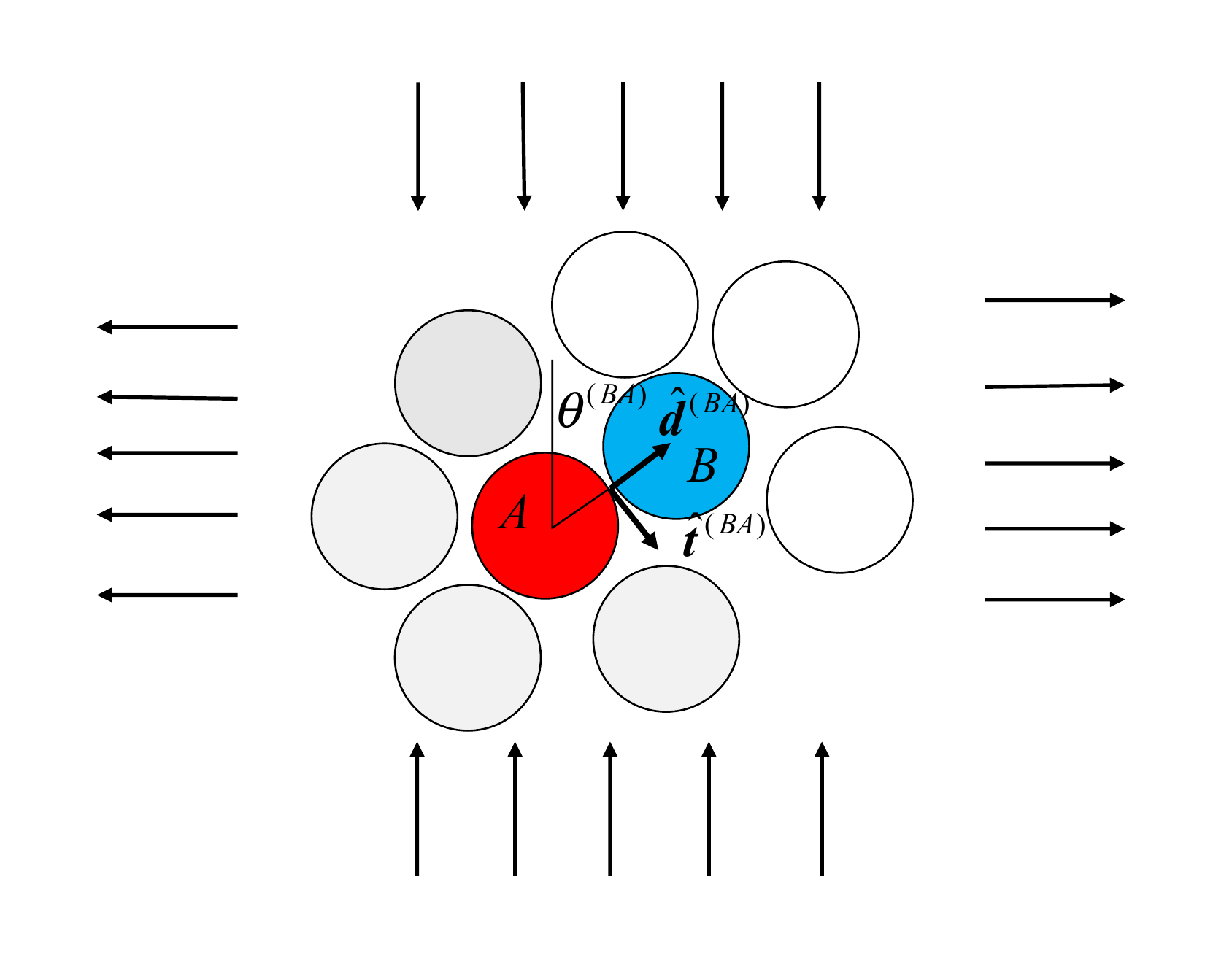}
  \caption{ The pair $AB$ and their near-contacting neighbours, 
with the angle $\theta^{(BA)}$ and the unit vectors $\boldsymbol{\hat{d}}^{(BA)}$ 
and $\boldsymbol{\hat{t}}^{(BA)}$ along and perpendicular to the line of centres, respectively. }
  \end{figure}
  
  \subsection{Kinematics}
  
In planar extensional flow, the relative motion of the centre of particle $B$
with respect to the centre of particle $A$ is 
\begin{equation}
v_{\alpha }^{(BA)}=\frac{{ds}^{(BA)}}{dt}\hat{d}_{\alpha }^{(BA)}
+2a\frac{d{\theta}^{(BA)}}{dt}\hat{t}_{\alpha }^{(BA)},
\end{equation}
where $s$ is the separation of the edges along the line of centres. 
The relative velocity of their points of near contact is then, 
\begin{equation}
  v_{\alpha }^{(BA)}+a(\omega ^{(A)}+\omega^{(B)}) \hat{t}_{\alpha }^{(BA)}
  \equiv v_{\alpha}^{(BA)}+aS \hat{t}_{\alpha }^{(BA)},
\end{equation}%
where $\omega$ is the angular velocity of the sphere and $S$ is their sum. 

The interaction of $A$ with a near-contacting neighbours $n$, other than $B$, is 
treated differently; the sphere $ n $ is assumed to move relative to $A$ with the average flow. 
Then, neglecting fluctuations in translational velocity, the
relative velocity of centres of pair $nA$ is 
\begin{equation}
v_{\alpha}^{(nA)}=2a \te{D}_{\alpha \beta}\hat{d}_{\beta }^{(nA)}
\end{equation}%
and the relative velocity of the points of near contact $nA$ is 
\begin{equation}
v_{\alpha}^{(nA)}+ a\omega^{(A)}\hat{t}_{\alpha }^{(nA)}.
\end{equation}

\subsection{Force}

The force of interaction between a typical pair $AB$ of particles is related
to the relative velocity and distance between their points of near
contact. According to Jeffrey and Onishi (1984) and Jeffrey (1992), the
force $\boldsymbol{F}^{(BA)}$ exerted by sphere $B$ on sphere $A$ through a fluid with viscosity $\mu$, is 
\begin{eqnarray}
  F_{\alpha }^{(BA)}  &=& 6\pi \mu a \te{K}_{\alpha \beta }^{(BA)} v_{\beta }^{(BA) }
  -\frac{F_{0}}{s^{(BA)}}\hat{d}_{\alpha}^{( BA)} -9.54\pi \mu a^{2} 
\left( \hat{t}_{\beta} \te{D}_{\beta \xi } \hat{d}_{\xi }\right) \hat{t}_{\alpha }^{(BA) }
\label{F_BA} \\
&& \phantom{} +\pi \mu a^{2}\left[ \ln \left( \frac{a}{s^{(BA) }}\right) -0.96\right] 
\omega^{(A)} \hat{t}_{\alpha}^{(BA)}+\pi \mu a^{2}\ln \left( \frac{a}{s^{(BA)}} \right) 
\omega^{(B)} \hat{t}_{\alpha}^{(BA)} \notag,
\end{eqnarray}
where
\begin{equation}
  \te{K}_{\alpha \beta }^{(BA)}=\frac{1}{4}\frac{a}{s^{(BA)}} \hat{d}_{\alpha}^{(BA)}
  \hat{d}_{\beta}^{(BA)}+\left[ \frac{1}{6}\ln \left( \frac{a}{s^{(BA)}}\right) +0.64 \right] 
  \hat{t}_{a}^{(BA)} \hat{t}_{\beta}^{(BA)}
\end{equation}%
and the constant terms have been retained because they are of a similar size, unless $s$ is extremely small. The interaction force also includes a short-range
repulsion of strength $F_{0}$ force times length (e.g., Singh and Nott, 2000).

We take the near-contacting neighbours, $m\neq B$, to be those that most influence equilibrium and make the greatest contribution to the stress. There are $k-1$ of these per sphere and we assume that the separation between their edges is $\bar{s}$. The number, $k$, of near-contacting neighbours is expected to be less, perhaps far less, than the number of nearest neighbours and to depend upon the area fraction, or concentration, $c$.

For the near-contacting neighbours, $m$, the corresponding force is based on the average motion and the separation $\bar{s}$: 
\begin{eqnarray}
F_{\alpha }^{(mA) } &=& \frac{3}{\bar{s}}a^{3} \pi \mu \left( 
\te{D}_{\beta \xi} \hat{d}_{\xi}^{(mA)} \hat{d}_{\beta}^{(mA)}\right) 
\hat{d}_{\alpha }^{(mA)}+\pi \mu a^{2}\left[\ln \left( \frac{a}{\bar{s}}\right) -0.96\right] 
\omega^{(A)} \hat{t}_{\alpha}^{(mA)}  \notag \\
&& \phantom{}  +  2 a^{2}\pi \mu \left[ \ln \left( \frac{a}{\bar{s}}\right) -0.96\right]
\left( \te{D}_{\beta \xi}\hat{t}_{\xi}^{(mA)} \hat{d}_{\beta}^{(mA)}\right) 
\hat{t}_{\alpha }^{(mA)}-\frac{F_{0}}{\bar{s}}\hat{d}_{\alpha}^{(mA)}.
\end{eqnarray}

\subsection{Force and Torque Balances}

In the more complicated two-dimensional simple shear flow, Jenkins and La Ragione (2015) require the balance of forces for a typical pair of spheres under the action of their near-contacting neighbours. Here, for the less complicated planar extensional flow, we consider the balance of both force and torque. The focus on a flow in which there is no average rotation makes this easier to do; and the possibility of solving for both the translational and rotational degrees of freedom of a typical pair should increase the accuracy of the modeling.

The balance of forces for particle $A$ is%
\begin{equation}
  F_{\alpha}^{(BA)} + \sum_{m\neq B}^{N^{(A)}}F_{\alpha}^{(mA)}=0;  \label{EqA}
\end{equation}
and that for particle $B$ is%
\begin{equation}
  F_{\alpha}^{(AB)} + \sum_{m\neq A}^{N^{(B)}}F_{\alpha}^{(mB)}=0 , \label{EqB}
\end{equation}%
with $F_{\alpha }^{(BA) }=-F_{\alpha }^{(AB)}$. 
The difference in the force balances projected along 
$\boldsymbol{\hat{d}}^{(BA) }$ is
\begin{equation}
3\pi \mu a\frac{a}{s^{(BA) }} \frac{ds^{(BA) }}{dt} -2\frac{%
F_{0}}{s^{(BA) }}+6\pi \mu a^{2}\frac{a}{\bar{s}}\hat{d}_{\alpha
}^{(BA) } \te{J}_{\alpha \beta \gamma }\te{D}_{\beta \gamma }-
2\frac{F_{0}}{\bar{s}} Y_{\alpha } \hat{d}_{\alpha }^{(BA) }=0;  \label{Norm_comp}
\end{equation}%
while along $\hat{t}_{\alpha }^{(BA) }$ it is 
\begin{eqnarray}
0 &=&4\left[ \ln \left( \frac{a}{s^{(BA)}}\right) +3.84\right] 
\frac{d\theta^{(BA)}}{dt}-19\hat{t}_{\beta }^{(BA) }\te{D}_{\beta \xi }
\hat{d}_{\xi}^{(BA) }  \label{Tang_comp} \\
&&+\left[ 2\ln \left( \frac{a}{s^{(BA)}}\right) -0.96\right] S
+\left[ \ln \left( \frac{a}{\bar{s}}\right) -0.96\right] S
\varepsilon_{\alpha \beta} Y_{\beta}^{(BA)} t_{\alpha}^{(BA)} \notag \\
&& + 6\frac{a}{\bar{s}} \te{D}_{\beta \xi }\te{J}_{\alpha \xi \beta}^{(BA)}
t_{\alpha}^{(BA)} + 2 \left[ 2\ln \left( \frac{a}{\bar{s}}\right) -1.92 \right] 
\te{D}_{\beta \xi }\te{J}_{\alpha \xi \beta }^{(BA)}t_{\alpha }^{(BA) } , \notag
\end{eqnarray}
with
\begin{equation}
\te{J}_{\alpha \xi \beta }^{(BA)} = \sum_{m\neq B}^{N^{(A) }}\hat{d}%
_{\alpha }^{(mA) }\hat{d}_{\beta }^{(mA) }\hat{d}%
_{\xi }^{(mA) }
\end{equation}%
and
\begin{equation}
  Y_{\alpha }^{(BA)} = \sum_{m\neq B}^{N^{(A) }}\hat{d}  _{\alpha }^{(mA) }.
\end{equation}%
In writing Eqs.~\eqref{Norm_comp} and \eqref{Tang_comp}, 
we assume that $\te{J}_{\alpha \xi \beta }^{(BA)}=- \te{J}_{\alpha \xi \beta }^{(AB)}$ and 
$Y_{\alpha }^{(BA) }=-Y_{\alpha }^{(AB) }$; that is, 
the arrangement of near-contacting neighbours of $B$ is the reflection of that of $A$.
The terms proportional to $S$ incorporate the influence of the rotations on the force balance.

The balance of torques for particle $A$ is 
\begin{equation}
\varepsilon _{\alpha \beta}d^{(BA)}_{\alpha}F_{\beta}^{(BA) }+\varepsilon
_{\alpha \beta}\sum_{m\neq B}^{N^{(A) }}d^{(mA)}_{\alpha}F_{\beta }^{(mA)}=0,
\end{equation}%
and that for particle $B$ is%
\begin{equation}
\varepsilon _{\alpha \beta }d^{(AB)}_{\alpha}F_{\beta }^{(AB) }+\varepsilon
_{\alpha \beta }\sum_{m\neq A}^{N^{\left( B\right) }}d^{(mB)}_{\alpha}F_{\beta }^{(mB)}=0;
\end{equation}
so their sum is
\begin{eqnarray}
0 &=&4\left[ \ln \left( \frac{a}{s^{(BA)}}\right) +3.84\right]
\frac{d\theta^{(BA)}}{dt}-19\hat{t}_{\mu}^{(BA)} \te{D}_{\mu \xi} \hat{d}_{\xi}^{(BA)} \\
&&+\left[ 2\ln \left( \frac{a}{s^{(BA)}}\right) -0.96\right] S+\left[ \ln
\left( \frac{a}{\bar{s}}\right) -0.96\right] S\left( k-1\right)  \notag \\
&&+2\left[ 2\ln \left( \frac{a}{\bar{s}}\right) -1.92\right] \varepsilon
_{\xi \nu } \te{A}_{\nu \mu }^{(BA) } \te{D}_{\mu \xi } , \notag
\end{eqnarray}
with
\begin{equation}
\te{A}_{\nu \mu }^{(BA)} = \sum_{m\neq B}^{N^{(A)}}
\hat{d}_{\nu }^{(mA)} \hat{d}_{\mu}^{(mA)}
\end{equation}%
and, again, $\te{A}_{\nu \mu }^{(BA) } = \te{A}_{\nu \mu }^{(AB)}$.

%The tensors
$\tens{A}$, $\tens{J}$ and $\boldsymbol{Y}$ provide information on the distribution of 
spheres in the plane about a typical pair $AB$, 
as shown in Fig.1. We assume here that the distributions 
about a pair at a given orientation is the average over all pairs at that orientation. 
These average distributions should depend on both $\boldsymbol{\hat{d}}^{(AB)}$ and $\tens{D}$. 
As do Jenkins and La Ragione (2015), we treat the local equilibrium 
with the approximation that $\tens{A}$, $\tens{J}$ and $\boldsymbol{Y}$ 
are independent of $\tens{D}$. Then,
\begin{equation}
A_{\nu \mu }^{(BA) }=b_{1}\delta _{\nu \mu }+b_{2}\hat{d}_{\mu
}^{(BA) }\hat{d}_{\nu }^{(BA) },
\end{equation}%
\begin{equation}
J_{\alpha \xi \beta }^{(BA)}=b_{3}\hat{d}_{\alpha }^{(BA)} \hat{d}_{\xi}^{(BA) }\hat{d}_{\beta }^{(BA) }+b_{4}\left(\hat{d}_{\alpha}^{(BA) } \delta_{\xi \beta}+\hat{d}_{\xi}^{(BA)}
\delta_{\alpha \beta} + \hat{d}_{\beta}^{(BA)} \delta _{\xi \alpha}\right),
\end{equation}%
and
\begin{equation}
Y_{\alpha }^{(BA)} = b_{5}\hat{d}_{\alpha }^{(BA)}
\end{equation}%

To calculate the coefficients, Jenkins et al. (2005) assume that given sphere $B$, 
the remaining near-contacting neighbours of $A$ are 
distributed uniformly around its circumference. 
The results are given as a function of coordination number $k$ through
\begin{equation}
  b=-\frac{3\sqrt{3}\left(k-1\right) }{16\pi },
\end{equation}%
by
\begin{equation}
  b_{1}=\frac{k-1}{2}-b,~b_{2}=2b,~b_{3}=0,\text{ and } b_{4}=b,~b_{5}=4b.
\label{bee}
\end{equation}

In the planar extensional flow of interest,%
\begin{equation}
  \hat{t}_{\mu }^{(BA)} \te{D}_{\mu \xi}\hat{d}_{\xi}^{(BA)} = \dot{\gamma}\sin 2\theta 
  \quad \text{ and } \quad
  \hat{d}_{\mu }^{(BA)} \te{D}_{\mu \xi}\hat{d}_{\xi}^{(BA)}=-\dot{\gamma}\cos 2\theta .
\end{equation}%

We use these in the differences of the components of the force balances, make lengths dimensionless by the sphere radius $a$, time by the inverse of the shear rate, forces by $\pi a^{2}\mu \dot{\gamma}$, write the
dimensionless strength of the repulsion as $\hat{F}=F_{0}/(\pi a^{3}\mu \dot{\gamma})$ 
and remove the superscript $(BA)$. Then, the normal
component becomes%
\begin{equation}
\frac{1}{s}\frac{ds}{d\gamma }=\frac{2}{3}\hat{F}\left( \frac{1}{s}+\frac{4b}{\bar{s}}\right) 
+\frac{4b}{\bar{s}}\cos 2\theta; \label{norm_eq}
\end{equation}%
and the tangential component is
\begin{equation}
  \left[ \ln \left( \frac{1}{s}\right) +3.84\right] \frac{d\theta }{d\gamma }= 
  \left[ c_{1}+c_{2}\ln \left( \frac{1}{s}\right) \right] \sin 2\theta ,
  \label{tang}
\end{equation}%
with 
\begin{equation}
c_{1}=4.77-3\frac{b}{\bar{s}}
\quad \text{ and } \quad 
c_{2}=\frac{6b}{(4b-k+1) } 
\frac{1/\bar{s}}{\ln \left( 1/\bar{s}\right) -0.96},
\label{cees}
\end{equation}%
and we have employed the difference in the force balances and the sum of the torque
balances to write
\begin{equation}
S=-2c_{2}\sin 2\theta . \label{S}
\end{equation}

The balances of force and torque, Eqs.\eqref{norm_eq}, \eqref{tang} and \eqref{S}, 
employed in Eq. \eqref{F_BA}, provide an expression for $\boldsymbol{F}^{(BA)}$ 
in terms of average quantities
\begin{align}
F_{\alpha }^{(BA) } =&4b\frac{F_{0}}{\bar{s}}\hat{d}_{\alpha}^{(BA)}
+\pi \mu a^{3}\frac{6b}{\bar{s}}\cos 2\theta \dot{\gamma} \hat{d}_{\alpha}^{(BA) }
+\pi \mu a^{2}\left(2c_{1}+0.96c_{2}\right) \sin 2\theta \dot{\gamma}\hat{t}_{a}^{(BA)}  \notag \\
&-9.54\pi \mu a^{2}\sin 2\theta \dot{\gamma}\hat{t}_{\alpha}^{(BA)}.  \label{final_F}
\end{align}
This is later used in the calculation of the stress.

\subsection{Representative trajectory}

The representative trajectory is a single trajectory that incorporates the influence of those that contribute most to the stress. Along the representative trajectory particle $B$ moves with respect to particle $A$ in a succession of states in which force and torque are balanced. The other near-contacting  particles, $m$, of the pair are assumed to move with the average flow, at the constant distance $\bar{s}$ from the pair. The equation that determines this trajectory results from the balances of force and torque and is a function of two parameters: the average number of near-contacting particles, $k$, and the distance, $\bar{s}$. Upon combining Eqs. \eqref{norm_eq} and \eqref{tang}, it is
\begin{equation}
\frac{ds}{d\theta }=\frac{2}{3\bar{s}}\frac{\hat{F}\left( \bar{s}+4bs\right)
+6bs\cos 2\theta }{\left[ c_{1}+c_{2}\ln \left( 1/s\right) \right] \sin
2\theta }\left[ \ln \left( \frac{1}{s}\right) +3.84\right] .  \label{ds_dth}
\end{equation}
Within the $\theta$ interval 0 to $\pi/2$, the trajectory begins at $\theta_{0}$ 
and ends at $\theta_{1}$, 
and both angles must be determined. Because of the presence of $\hat{F}$, the trajectory is 
asymmetric about $\pi/4$, and $\theta_{0}$  differs from $\pi/2-\theta_{1}$. 

The amount of total 
strain, $\hat{\gamma}$, necessary to complete the trajectory may be calculated from the pair interaction 
in the average flow. From Eq. \eqref{tang}
\begin{equation}
\frac{d\gamma }{d\theta }=\frac{\ln \left( 1/s\right) +3.84}{\left[
c_{1}+c_{2}\ln \left( 1/s\right) \right] \sin 2\theta } \label{dg};
\end{equation}
so,
\begin{equation}
\hat{\gamma}=\int_{\theta _{0}}^{\theta _{1}}\frac{\ln \left( 1/s\right)
+3.84}{\left[ c_{1}+c_{2}\ln \left( 1/s\right) \right] \sin 2\theta }d\theta.
 \label{gamma}
\end{equation}%

\subsection{Particle distribution}

We next introduce the distribution of near-contacting neighbours along the trajectory, $A(\theta)$, defined so that  $A( \theta ) d\theta $ is the average number of such particles within the element $d\theta$. At steady state, the flux, $A ( \theta )  d\theta /d\gamma$, of these
equilibrated particles along the trajectory is constant. That is, particles are more likely to be 
where the velocity along the trajectory is least. Because the repulsive force breaks the symmetry 
of approach and departure, the distribution is anticipated to be asymmetric about $\pi /4$. 
In computations, we implement the flux condition as a differential equation 
\begin{equation}
\frac{dA}{d\theta }=-\frac{A}{\dot{\theta}}\frac{d\dot{\theta}}{d\theta },
\label{d_A}
\end{equation}%
with%
\begin{equation}
\frac{d\dot{\theta}}{d\theta }=\frac{\partial \dot{\theta}}{\partial \theta }%
+\frac{\partial \dot{\theta}}{\partial s}\frac{ds}{d\theta }.
\end{equation}%

The distribution $A( \theta ) $ is related to the average number 
near-contacting neighbours per particle by
\begin{equation}
4\int_{\theta_{0}}^{\theta_{1}} A( \theta ) d\theta = k.
\end{equation}%
We implement this as a differential equation for the partial number of near-contacting neighbours %
\begin{equation}
 I( \theta ) \equiv \int_{\theta_{0}}^{\theta }A ( \theta^{\prime }) 
d\theta^{\prime},
\end{equation}
as
\begin{equation}
\frac{dI}{d\theta }= A ( \theta )   \label{d_I},
\end{equation}%
with boundary conditions $I ( \theta_{0}) =0$ and $I(\theta_{1})=k/4$. 

Given that the beginning and ending angles of the trajectory differ, we take the beginning and ending values of the particle separation to be the same. There are three first-order differential equations, Eqs. \eqref{ds_dth}, \eqref{d_A} and \eqref{d_I}, for $s$, $A$ and $I$ as functions of $\theta$, and four boundary conditions: one for each of $s_{0}$ and $s_{1} $ that introduce a single parameter, and two for $I$. Consequently, $\theta_{1}$ may be determined as part of the solution. The inputs are $\theta_0$, $s_0 = s_1$, $\bar{s}$ and $k$.  In Appendix B, we provide the Matlab code that is employed in the solver.  We generate solutions and compare them with the results of Stokesian dynamics simulations in a later section.

\section{Particle stress}

Knowledge of the distribution of near-contacting neighbours $A( \theta )$ and 
the contact forces along the trajectory permits the calculation of the
macroscopic particle stress in the suspension. The stress tensor is, according to
Cauchy (Love 1944, Appendix, Note B),%
\begin{equation}
\te{T}_{\alpha \beta }=na\int_{0}^{2\pi }A (\theta) F_{\alpha }\hat{d}_{\beta}d\theta,
\label{str}
\end{equation}%
where $n$ is the number of particles per unit area and $F_{\alpha}$ is given by Eq.\,\eqref{final_F}. 
Then, the two-dimensional viscosity is $2a\mu$. The dimensionless form, 
$\te{t}_{\alpha\beta } = \te{T}_{\alpha\beta}/(2a\mu\dot{\gamma})$,
with $n=c/(\pi a^{2})$, is
\begin{eqnarray}
\te{t}_{\alpha \beta } &=& c  \frac{b}{\bar{s}}\int_{0}^{2\pi }A\left( \theta \right)
\left( 2\hat{F}+3\cos 2\theta \right) \hat{d}_{\alpha }\hat{d}_{\beta} d\theta \notag \\
&&+ c  \left( c_{1}+0.48c_{2}-4.77\right) \int_{0}^{2\pi }A\left( \theta \right) 
\sin 2\theta \hat{t}_{a}\hat{d}_{\beta }d\theta. 
\end{eqnarray}%
The particle shear stress,%
\begin{equation}
\tau \equiv \frac{1}{2}\left( \te{t}_{11}-\te{t}_{22}\right), 
\end{equation}%
is%
\begin{eqnarray}
\tau&=&-2c\frac{b}{\bar{s}}\int_{0}^{\pi/2 }A\left( \theta \right)
\left( 2\hat{F}+3\cos 2\theta \right) \cos 2\theta d\theta  \notag \\
&&+c\left( 2c_{1}+0.96c_{2}-9.54\right) \int_{0}^{\pi/2 }A\left( \theta
\right) \sin^{2} 2\theta d\theta,
\end{eqnarray}%
where $b$, $c_1$ and $c_2$ are given in terms of $k$ in  Eqs. \eqref{bee} and \eqref{cees}, respectively.
The shear stress depends on the separation, $\bar{s}$, of near-contacting neighbours other than $B$, and on the area fraction, explicitly and through the coordination number, $k$. Because the direct contribution of the repulsive force to the integral is very small and the trigonometric factors associated with the other contributions are even about $\pi/4$, the shear stress is independent of the asymmetry of the particle distribution about $\pi/4$. In contrast, this asymmetry is crucial to the determination of the particle pressure.

The particle pressure,
\begin{equation}
p\equiv -\frac{1}{2}\left( t_{11}+t_{22}\right) ,
\end{equation}%
is%
\begin{equation}
  p=-2c\frac{b}{\bar{s}}\int_{0}^{\pi/2 }A\left( \theta \right) 
  \left( 2\hat{F}+3\cos 2\theta \right) d\theta .
\end{equation}%
This pressure also depends on $\bar{s}$ and $c$ and its existence is due to the asymmetry of $A$ about $\pi/4$. This asymmetry is due to that of the separation along the representative trajectory created by $\hat{F}$  and the influence of the asymmetry of the separation 
on the angular velocity, $\dot{\theta}$. The particle pressure and the mechanisms responsible for it are a focus of this paper; a particle shear stress may be calculated based on the average flow, although that determined here is several times less than this, because of the approximate satisfaction of equilibrium.

Particle stresses associated with motion along the representative trajectories are compared with those measured in Stokesian dynamics simulations after a discussion of the simulations.

\section{Stokesian Dynamics}

We determine the trajectories of spherical particles in the flows by performing simulation with the same conditions as the theory (a monolayer with no inertia). We impose a planar extensional flow with shear rate $\dot{\gamma}$,
\footnote{Note: This is equivalent to the extensional rate $\dot{\varepsilon}$ in Seto et al., 2017.}
\begin{equation}
 \boldsymbol{u}^{\infty} (\boldsymbol{r}) = \tens{D} \cdot \boldsymbol{r},
\quad
\tens{D}
=
\begin{pmatrix}
 \dot{\gamma} & 0 \\
 0 & - \dot{\gamma}
\end{pmatrix}.
\end{equation}
A simulation box with periodic boundary conditions constantly deforms according to this velocity gradient $\tens{D}$. Significant deformations of the simulation box can be avoided by using the Kraynik--Reinelt periodic boundary conditions, which rearrange the deformed box to the original square box after a constant strain interval (Kraynik and Reinelt 1992, Todd and Daivis, 1998, Seto et al., 2017). Thus, the flow can be applied for a sufficiently long time to evaluate its steady states. 

Due to the negligible inertia of the particles, translational and angular velocities can be determined by solving the force and torque balance equations for the respective particles ($i = 1, \dotsc, N$):
\begin{equation}
\begin{pmatrix}
\boldsymbol{0} \\
\boldsymbol{0}
\end{pmatrix}
=
\begin{pmatrix}
\boldsymbol{F}_{\mathrm{H}} \\
\boldsymbol{T}_{\mathrm{H}} 
\end{pmatrix}
+
\begin{pmatrix}
\boldsymbol{F}_{\mathrm{R}} \\
\boldsymbol{0}
\end{pmatrix}.\label{112336_10Apr20}
\end{equation}
Here, a vector, such as $\boldsymbol{F}_{\mathrm{H}}$, represents all $N$ particles, $\boldsymbol{F}_{\mathrm{H}} \equiv (\boldsymbol{F}_{\mathrm{H}}^{(1)}, \dotsc, \boldsymbol{F}_{\mathrm{H}}^{(N)} )$.

The hydrodynamic interactions in the Stokes, zero Reynolds number, regime are linear in the velocities
\begin{equation}
\begin{pmatrix}
\boldsymbol{F}_{\mathrm{H}} \\
\boldsymbol{T}_{\mathrm{H}} 
\end{pmatrix}
= 
- 
\tens{R}_{\mathrm{FU}}
\cdot
\begin{pmatrix}
\boldsymbol{U}- \boldsymbol{u}^{\infty} \\
\boldsymbol{\Omega} 
\end{pmatrix}
+
\tens{R}_{\mathrm{FE}}:\tens{D}_{\mathrm{N}},\label{112626_10Apr20}
\end{equation}
where $\tens{D}_{\mathrm{N}}$ is block diagonal of $N$ copies of $\tens{D}$. There exist several levels of approximations to construct the resistance matrices 
$\tens{R}_{\mathrm{FU}}$ and $\tens{R}_{\mathrm{FD}}$.
Brady and Bossis (1988) constructed them using truncated multipole expansions for the far-field
interactions and a pairwise solution for lubrication interactions. In this work, 
we focus on a special situation in which repulsive forces are very weak in comparison with viscous drag forces. Under such conditions, particles tend to approach their neighbours very closely. 

Because the resistance coefficients diverge at contacts ($s = 0$), the nearly touching hydrodynamic interactions dominate the dynamics. This is why we  construct the approximate resistance matrices 
with the leading $1/s$ term in the normal component and the logarithmic term $\log (1/s)$ and following constants  in the tangential component, using the solution for two nearly touching rigid spheres (Jeffrey and Onishi, 1984, Jeffrey, 1992).
(A detailed description can be seen elsewhere, c.f. Mari el al., 2014.)
The hydrodynamic interaction is effective only when $s < 0.10$; thus, the resistance coefficients remain positive in this range.

The repulsive force employed in this work is the same as that used in Nott and Brady (1994):
\begin{equation}
  \boldsymbol{F}_{\mathrm{R}} = 
  F_0 \frac{\lambda^{-1} e^{- s/\lambda}}{1-e^{- s/\lambda}}
  \boldsymbol{n},
  \label{112629_10Apr20}
\end{equation}
where the range of repulsive force is set by a parameter $\lambda$.
Because the repulsive force diverges as $F_0/s$ in the limit of contact, $s \to 0$, 
some force balance can occur at a finite gap.
Because the repulsive force diverges as $F_0/s$ in the limit of contact, $s \to 0$, 
some force balance can occur at a finite gap.
Thus, the gap $s$ remains positive, 
and contact forces do not appear in the current system. 
Note that the divergence in the lubrication coefficient
does not guarantee the presence of a minimum $s >0$,
thus it leads to a pathologic singularity in theoretical models (Ball and Melrose 1995).
%

%%%%%%%%%%%%%%%%%%%%%%%%%%%%%%%%%%%%%%%%%%%%%%%%%%%%%%%%%%%%%%%%%%%%%%%%%%%%%%%%%%%%%%%%%%%%%%%%%%%%

By solving the force and torque balance equations \eqref{112336_10Apr20}
with the hydrodynamic interaction \eqref{112626_10Apr20} and repulsive force \eqref{112629_10Apr20},
the linear and angular velocities $(\boldsymbol{U}, \boldsymbol{\Omega})$
can be determined at each time step. Integrating these velocities $\boldsymbol{U} $
with a discretized time step, we obtain trajectories of particles. 
The particle stress tensor $\hat{\tens{T}}$ is given by the symmetrized first moment
\begin{equation}
\hat{\tens{T}}
=
\frac{1}{V}
\sum_{j}
\frac{
 \boldsymbol{r}^{ij} \boldsymbol{F}^{ij} + \boldsymbol{r}^{ji} \boldsymbol{F}^{ji}}{2},
\label{160608_13Apr20}
\end{equation}
with the pairwise forces 
$\boldsymbol{F}^{ij} \equiv \boldsymbol{F}^{ij}_{\mathrm{Lub}}+\boldsymbol{F}^{ij}_{\mathrm{R}}$ 
and relative positions $ \boldsymbol{r}^{ij} \equiv \boldsymbol{r}^{i} -\boldsymbol{r}^{j}$
of all interacting particle pairs. Here, $V = 2a L^2$ is the volume of the monolayer system. Normalizing the symmetrized first moment with the shear stress of the suspending fluid $2 \mu \dot{\gamma}$, gives the dimensionless stress $\te{t}_{\alpha \beta} \equiv \hat{\te{T}}_{\alpha \beta}/2 \mu \dot{\gamma}$.
Thus, we have the dimensionless particle pressure
 $p \equiv -(\te{t}_{11}+\te{t}_{22})/2 $
and the dimensionless particle shear stress
 $\tau \equiv (\te{t}_{11}-\te{t}_{22})/2 $,
respectively. 

\section{Results}

%%%%%%%%%%%%%%%%%%%%%%%%%%%%%%%%%%%%%%%%%%%%%%%%%%%%%%%%%%%%%%%%%%%%%%%%%%%%%%%%%%%%%%%%%%%%%%%%%%%%

We simulate monolayer systems with 1000 spheres of radius $a$
at area fractions, $c$, between $0.52$ and $0.64$.
We generate initial configurations 
with a simple algorithm using random numbers.
To reduce effects of such artificially generated initial configurations,
the post-processing analyses use steady state data from 10 to 50 strain units.
The repulsive force is set to be very weak $F_0/\pi a^3 \mu \dot{\gamma} = 10^{-4}$
and short-ranged $\lambda/a = 10^{-2}$.

In the implementation of the model, we take $\bar{s} = 0.02$, $\theta_0 = 10^{-6}$, $s_0 = s_1 = 0.10$ and assume that $k$ varies linearly with $c$ from 2.0, at $c = 0.52,$ to 2.5, at $c = 0.64$. These values and the relation for $k$ are plausible and they are influenced by those measured in the simulations. The value $\bar{s}$ gives values of the shear stresses that are close to those measured. The value of $\theta_0$ is the default absolute tolerance of the solver; smaller values of $\theta_0$ have little influence on the shear stress, but do slightly improve the prediction of the pressure. The initial and final values of the separation were those employed in the simulations, and the variation of near-contacting neighbours with concentration, $k(c)$, was that measured in the simulation.

\begin{figure}
\centering
\includegraphics[width=10cm]{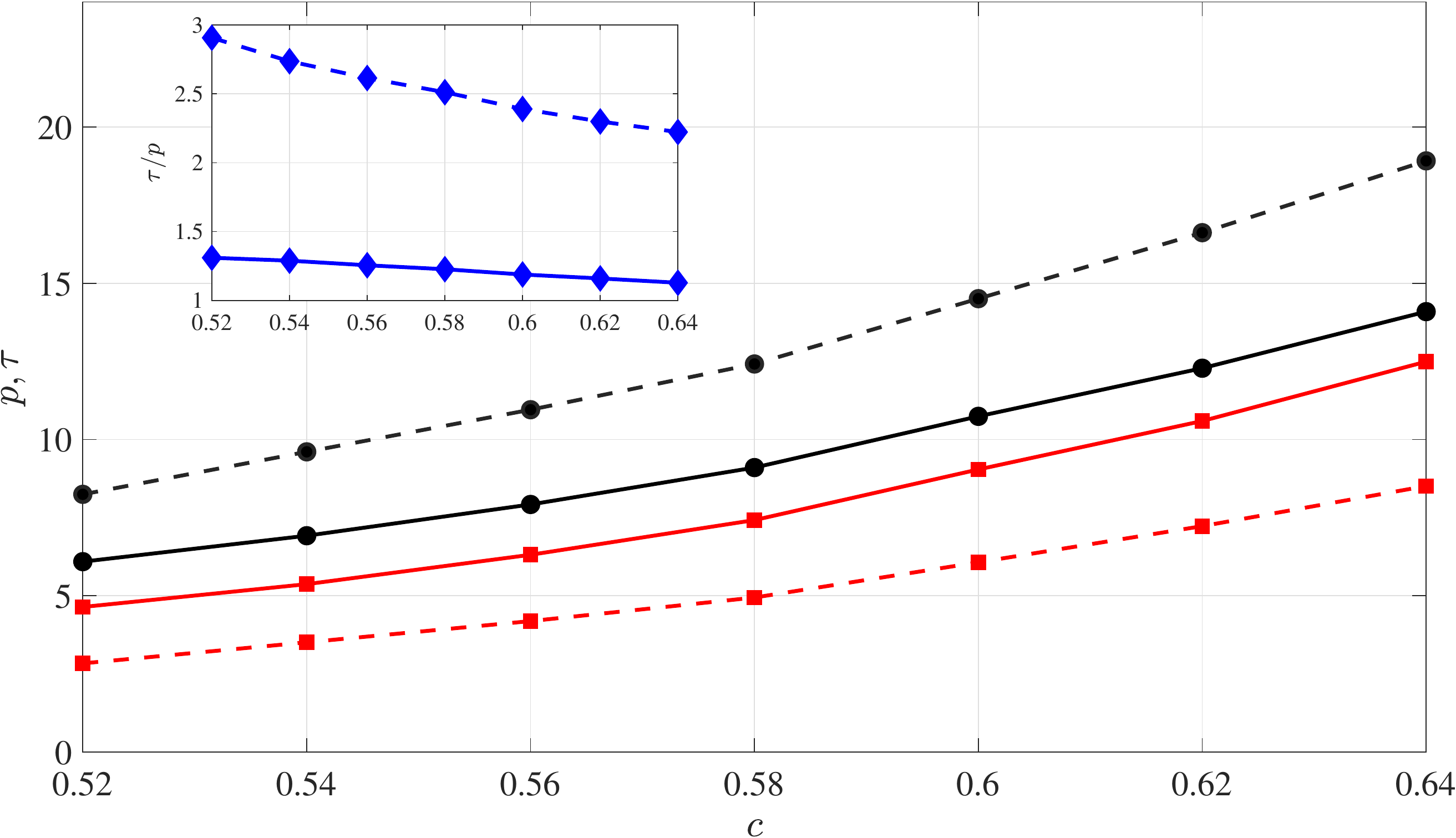} 
\caption{Shear stress $\tau$ (black), pressure $p$ (red) and stress ratio (blue) over a range of concentrations, as measured in the simulation (solid) and predicted by the model (dashed).}
\label{tau_p}
\end{figure}

Fig. $\ref{tau_p}$ shows plots of the shear stress $ \tau  $ and pressure $ p$ measured in the simulation and predicted by the model, over a range of area fraction $c.$ The stresses in the simulation increase in a similar manner with $c$,
but the ratio $\tau / p$ decreases gradually. The predicted particle pressure is somewhat less than that measured in the simulations and the predicted shear stress is somewhat greater. The ratio of shear stress to pressure decreases with area fraction, as in the numerical simulations; but, because of under- and over-predicting, we have a greater value for the ratio.

\begin{figure}
\centering
\includegraphics[width=13.5cm]{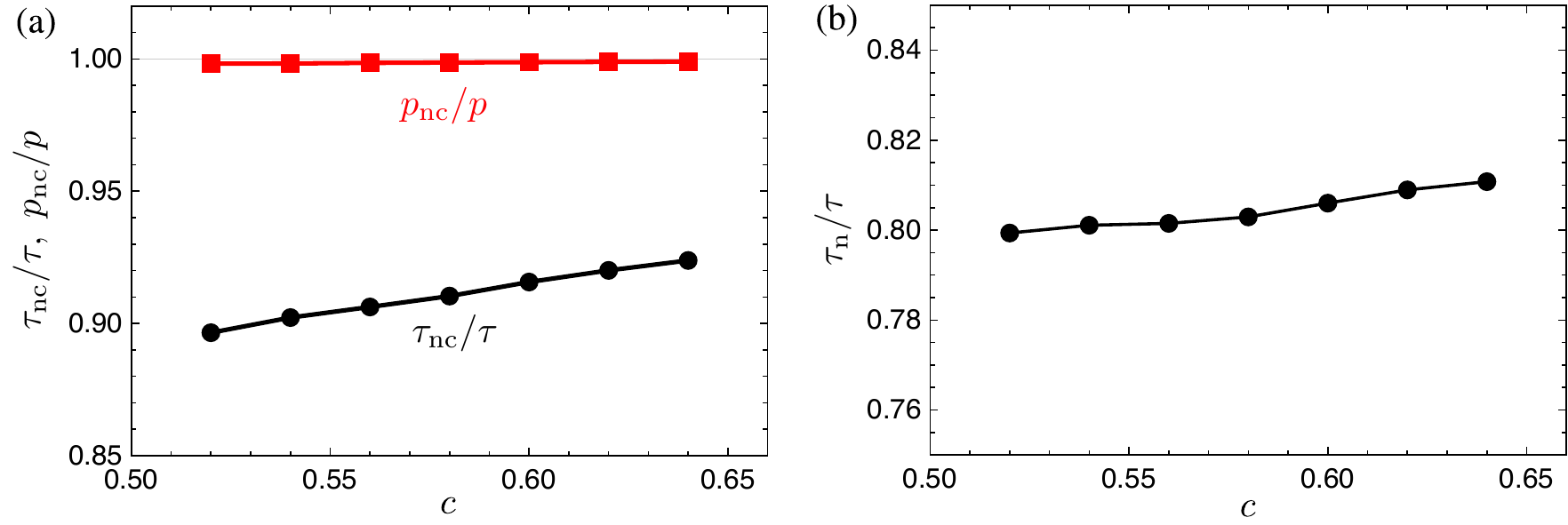} 
\caption{(a) Near-contact contributions to $\tau$ and $p$.
(b) The partial stress obtained using only the normal component of the pairwise force in
 \eqref{160608_13Apr20}.}
\label{contact}
\end{figure}

Most of stress measured in the simulations is generated by closely approaching particles - defined as those with a separation less than one per cent of the particle radius. As seen in Fig.~$\ref{contact}$, more than 90\% of shear stress is generated from particle pairs with $s < 0.01$. Moreover, such near-contacting particles generates almost 100\% of the pressure $p$. Finally, approximately 80 \% of the shear stress $\tau$ comes from the normal force.

\begin{figure}
\centering
\includegraphics[width=1.03\textwidth]{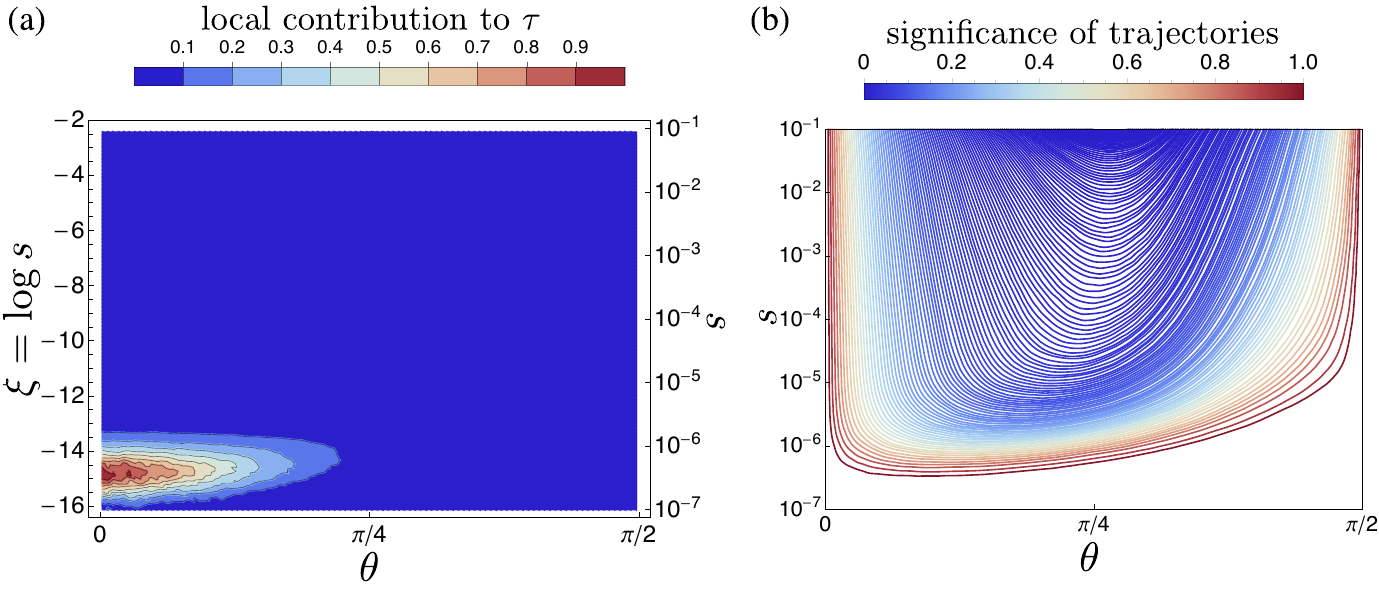} 
\caption{(a) Stress concentration, (b) trajectories, with colors denoting their average contribution to the stress. Both figures are for $c$ = 0.52.}
\label{Ryohei}
\end{figure}

We can check the concentration of stress contribution
in the very narrow range of $s$ using distribution maps.
We calculate 
the spatial distribution in $\xi \equiv \log s $.
The statistics are 
calculated with
discretized bins $\xi_{k} \equiv \xi_{1}+(k-1) \Delta \xi$,
$k = 1, \dotsc, k_{\mathrm{max}}$.
$\xi_{1} = \log{10^{-7}}$
and $\xi_{k_\mathrm{max}} = \log{10^{-1}}$.
The results are 
plotted with $s$ in a logarithmic scale.
Fig. $\ref{Ryohei} $(a) displays
the distribution of shear stress, indicating that the stress tends to be concentrated near the stagnation point $(\theta, s) \approx (0, 10^{-6})$. The region of concentration  spreads until $\theta \sim \pi/4$.
We also separately calculate the stress of \eqref{160608_13Apr20} constructed with  normal forces $\boldsymbol{F}^{ij}_{\mathrm{n }} \equiv \boldsymbol{F}^{ij}\cdot \boldsymbol{n}^{ij} \boldsymbol{n}^{ij}$
and tangential forces
$\boldsymbol{F}^{ij}_{\mathrm{t}}\equiv
\boldsymbol{F}^{ij}-\boldsymbol{F}^{ij}\cdot \boldsymbol{n}^{ij} \boldsymbol{n}^{ij}$.
As shown in Fig. $\ref{contact}$(b), 80 \% of the shear stress $\tau$ indeed comes from the normal forces.
%%%%%%%%%%%%%%%%%%%%%%%%%%%%%%%%%%%%%%%%%%%%%%%%%%%%%%%%%%%%%%%%%%%%%%%%%%%%%%%%%%%%%%%%%%%%%%%%%%%%
Besides systematic motions due to the shearing deformation,
particle motions fluctuate due to occasional configurations of surrounding particles.

Therefore,
it is necessary to reconstruct averaged trajectories
to compare with theoretical ones.
To this end, we first calculate the averaged 
relative-velocity field 
$\langle \boldsymbol{U}^{(j)} - \boldsymbol{U}^{(i)}\rangle$ 
over all interacting pairs $i$ and $j$
in terms of the relative position coordinate $\Delta \boldsymbol{r}^{ij}=(2a+s, \theta)$.
Owing to the symmetry of planar extension,
the statistics are taken on a quadrant: $0 < \theta < \pi/2$.
Because we consider a situation that is very close to the singularity (Ball and Melrose 1995),
the particles tend to approach very close to contact,
i.e., a bundle of trajectories is compressed into an extremely narrow range of $s$.
To avoid a loss of precision due to averaging, we carry out the statistical data binning 
with $\xi$ instead of $s$.

Once we evaluate the velocity field in the $\xi$--$\theta$ space,
i.e., $( \langle \dot{\xi} \rangle = \langle \dot{s}/s \rangle,  \langle  \dot{\theta} \rangle)$,
we can obtain trajectories as streamlines of the velocity field.
In Fig.\,$\ref{Ryohei}$(b), trajectories of the system with $c = 0.52$ and various initial positions are plotted. The trajectories are colored from blue to red, according to their contribution to the stress. We identify the band of significant trajectories as those with a separation of less than $10^{-2}$.

\begin{figure}
\centering
\includegraphics[width=0.65\textwidth]{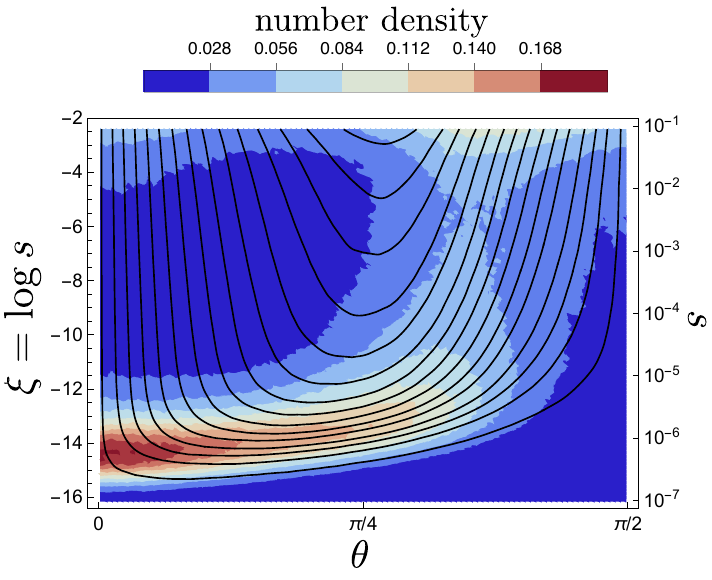} 
\caption{Particle number density over $s$ and $\theta$, with particle trajectories superposed for $c$ = 0.64.}
\label{Ryohei3}
\end{figure}

\begin{figure}
\centering
\begin{subfigure}
  \centering
  \includegraphics[width=.495\textwidth]{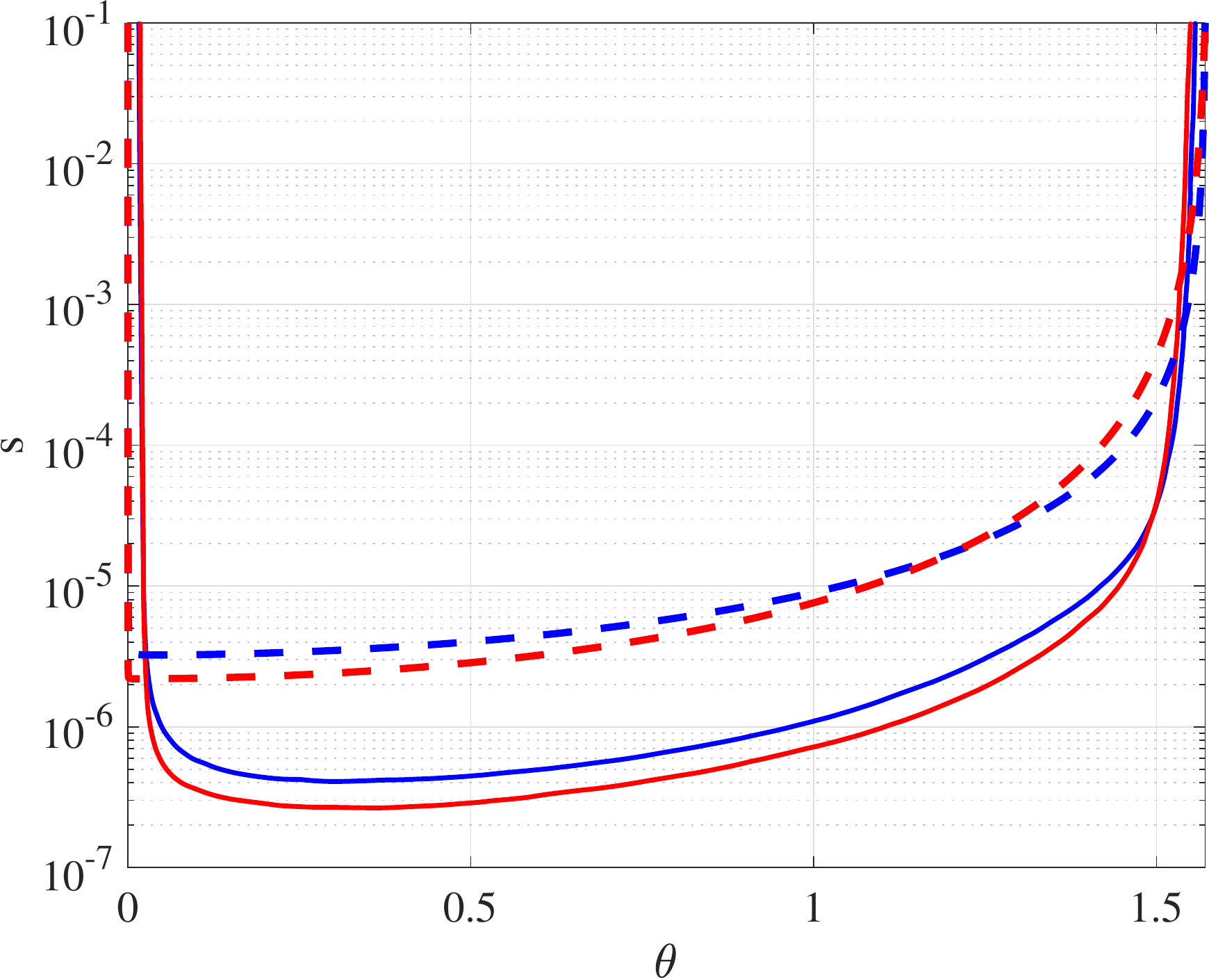}
  \label{fig:sub1}
\end{subfigure}\begin{subfigure}
  \centering
  \includegraphics[width=.495\textwidth]{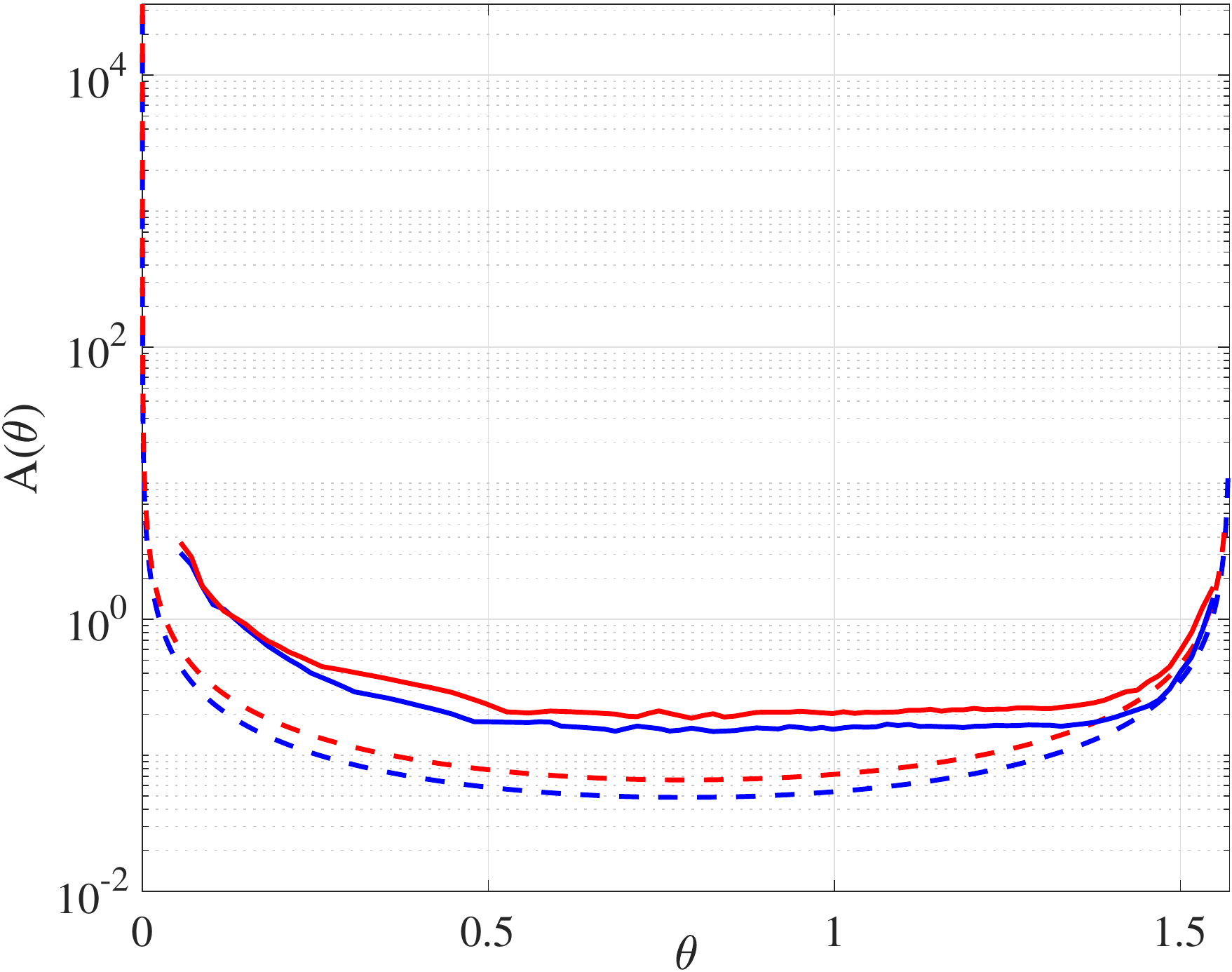}
  \label{fig:sub2}
\end{subfigure}
\caption{(a) Closest trajectories in the numerical simulation (solid) and the representative trajectories (dashed), for $c$ = 0.52 (blue) and $c$ = 0.64 (red). (b) The distribution $A(\theta)$ along the representative trajectory, as measured in the simulation (solid) and predicted by the model (dashed), for $c$ = 0.52 (blue) and $c$ = 0.64 (red).}
\label{fig:test}
\end{figure}

In Fig. 5, we show  the particle number density, measured in the simulations for $c = 0.64,$ with particle trajectories superposed. The near-contact distribution $A(\theta),$ normalized by the total number of particles, is obtained by integrating across the trajectories in the region below $s = 10^{-2}$. The product of the integral of the particle probability distribution over the area of Fig. 5 below $s = 10^{-2}$ and the total number of particles  is the coordination number.

In Fig. 6(a), we plot the trajectories from the simulation, for the smallest separation in Fig. 4(b), and the predicted representative trajectories of the model for concentrations of 0.52 and 0.64. The representative trajectories are located within the band and have a shape similar to the individual trajectories at smaller separations. In Fig. 6(b), we show distributions, $A(\theta)$, measured in the simulation and predicted along the representative trajectories for two values of the concentration. These share the same features and have a similar agreement as the trajectories. The asymmetry of the distributions result from the influence of the repulsive force that breaks the symmetry of approach and departure. 

We have employed information from the simulation on the variation in the coordination number as a function of concentration and the value of separation of near-contacting neighbours necessary to reproduce the measured particle shear stress. These, used in the model, gives it the capacity to generate particle trajectories that are representative of the stress-producing trajectories of the simulation and particle distributions along the representative trajectories with the appropriate asymmetry about $\pi/4$ to predict a reasonable variation of a particle pressure. .

We next indicate how the structure of the model can be used as the basis for a continuum theory of dense suspensions and to provide a context for existing theories.

\section{Tensorial formulation}

As elaborated upon by Onat and Leckie (1988) and Advani and Tucker (1987,1990), the distribution of near-contacting 
neighbours can be represented by an infinite series with respect to basis functions, such as
\begin{equation}
\te{f}_{\alpha \beta }=\hat{d}_{\alpha }\hat{d}_{\beta }-\frac{1}{2}\delta
_{\alpha \beta }
\end{equation}%
and
\begin{eqnarray}
\te{g}_{\xi \eta \rho \beta } &=&\hat{d}_{\xi }\hat{d}_{\eta }\hat{d}_{\rho }\hat{%
d}_{\beta }-\frac{1}{6}\left( \delta _{\xi \eta }\hat{d}_{\rho }\hat{d}%
_{\beta }+\hat{d}_{\xi }\hat{d}_{\eta }\delta _{\rho \beta }+\hat{d}_{\xi }%
\hat{d}_{\rho }\delta _{\eta \beta }+\hat{d}_{\eta }\hat{d}_{\beta }\delta
_{\xi \rho }+\delta _{\xi \beta }\hat{d}_{\eta }\hat{d}_{\rho }+\hat{d}_{\xi
}\hat{d}_{\beta }\delta _{\eta \rho }\right)   \notag \\
&&+\frac{1}{24}\left( \delta _{\xi \eta }\delta _{\rho \beta }+\delta _{\xi
\rho }\delta _{\eta \beta }+\delta _{\xi \beta }\delta _{\eta \rho }\right):
\label{g}
\end{eqnarray}
\begin{equation}
A(\theta )=\frac{k}{2\pi }\left( 1+4\mathcal{Z}_{\alpha \beta }f_{\alpha
\beta }+16\mathcal{B}_{\xi \eta \rho \beta }g_{\xi \eta \rho \beta
}+ \cdots \right).   \label{dist}
\end{equation}
The coefficients $\mathbfcal{Z}$ and $\mathbfcal{B}$ are completely traceless and completely symmetric tensors, related to the
distribution through
\begin{equation}
\mathcal{Z}_{\alpha \beta }=\int_{0}^{2\pi }A(\theta )\te{f}_{\alpha \beta
}d\theta  \label{Z}
\end{equation}%
and
\begin{equation}
\mathcal{B}_{\xi \eta \rho \beta }=\int_{0}^{2\pi }A(\theta )\te{g}_{\xi \eta
\rho \beta }d\theta .  \label{B_c}
\end{equation}%
These are the second and fourth moments of the distribution, respectively.

The stress of Eq.{\ref{str}) may be written in terms of these as
\begin{align}
\te{T}_{\alpha \beta } =&4 n a k\frac{F_{0}}{\bar{s}}b\delta _{a\beta }-
n\pi \mu a^{3}\left( M+N\right) \mathcal{B}_{\alpha \beta \gamma \eta } 
 \te{D}_{\gamma \eta }  \notag 
-n\pi \mu a^{3}\left( M-N\right)  
\frac{k}{4}\delta_{\alpha \eta} \delta_{\beta \gamma}\te{D}_{\gamma \eta }  \notag \\
&-n\pi \mu a^{3}\left(M+N\right) 
\frac{1}{6}\delta _{\alpha \beta }\mathcal{Z}_{\eta \gamma} \te{D}_{\gamma \eta }
\notag -n\pi \mu a^{3}\frac{2M-N}{6} \left( \delta_{\alpha \eta}\mathcal{Z}_{\gamma
\beta } +\delta_{\beta \eta}\mathcal{Z}_{\gamma
\alpha}\right)  \te{D}_{\eta \gamma },
\end{align}%
where
\begin{equation}
M=\frac{6b}{\bar{s}/a} \quad \text{and} \quad N=2c_{1}+0.96c_{2}-9.54;
\end{equation}
or, more compactly, as
\begin{equation}
\te{T}_{\alpha \beta }=na\left( 4kb\frac{F_{0}}{\bar{s}}
\delta_{a\beta} + \te{G}_{\alpha \beta \gamma \eta} \te{D}_{\gamma \eta }\right),
\label{constituv}
\end{equation}%
where
\begin{eqnarray}
\te{G}_{\alpha \beta \gamma \eta} &=&-n\pi \mu a^{3} \left[\frac{M-N}{4} 
k\delta_{\alpha \eta} \delta_{\beta \gamma} +\frac{M+N}{6} 
\delta _{\alpha \beta }\mathcal{Z}_{\eta \gamma }
+\frac{2M-N}{6} \left( \delta_{\alpha \eta}\mathcal{Z}_{\gamma
\beta } +\delta_{\beta \eta}\mathcal{Z}_{\gamma
\alpha}\right)\right]\nonumber\\
&-&n\pi \mu a^{3}(M+N) \mathcal{B}_{\alpha \beta \gamma \eta } .
\end{eqnarray}
The stress depends only on the second and fourth moments of the distribution, although an approximation of the distribution in terms of these does not provide a good representation of it. Because we predict the distribution function, we are able to capture the essential role played by the fourth moment. This places our theory in the context of the work of Chacko, et al. (2018), in which numerical simulations confirm the need of the fourth moment, in addition to the second moment, to describe stress reversal.

With knowledge of the distribution function, it is possible to evaluate the components of the second and fourth moments. In particular, when $F=0$, the only non-zero components are $\mathcal{B}_{1111}=\mathcal{B}_{2222}=-\mathcal{B}_{1122}$; when $F \ne 0$, then $\mathcal{Z}_{11}=-\mathcal{Z}_{22}$ are also different from zero. For example, when $c=0.64$, $k = 2.50$ and $\hat{F} = 10^{-4}$, their numerical values are  $\mathcal{B}_{1111} = 0.25$ and $\mathcal{Z}_{11} = -0.57$. Then, with Eq. \eqref{constituv}, the dimensionless particle pressure is
\begin{equation}
p=c \frac{b}{\bar{s}} \left(F-3\mathcal{Z}_{11} \right) = 8.51
\end{equation}

Given the force and torque balances, it is possible to characterize the role played by the torque balance in determining features of the trajectory and the distribution of particles along it. Ignoring the torque balance is equivalent to taking $S=0$, or $c_2=0$ in Eq.\,\eqref{tang}. This has an important influence on $\dot{\theta}$. Then, because both the distribution of near-contacting neighbours and the interval over which it is defined depend upon $\dot{\theta}$, there is a dependence of the stress upon it. For example, when $c=0.64$ and the torque balance is ignored, $p$ is $14.81,$ rather than $8.51$. That is, the value of $p$ is affected by the balance through the distribution. In contrast, $\tau$ is 18.60, rather than 18.94 and changes little, because it is independent of the shape of the distribution. 

In the absence of the knowledge of the distribution function, it is possible to develop evolution equations for the approximate determination of its moments (e.g., Prantil et al., 1993). Phan-Thien (1995) and Stickel et al. (2006) employ such an equation for the second moment, and break the symmetry of approach and departure by including a term in it that is proportional to $[\text{tr}(\tens{D}^{2})/2]^{1/2}$. Goddard (2006) introduces a memory integral for the second moment -- a representation for the solution to its evolution equation -- that breaks this symmetry by incorporating a term proportional to the $\tens{D}^{2}$. Gillissen and Wilson (2018, 2019) and Gilissen et al. (2019) distinguish between an anisotropic distribution of near contacts and an isotropic distribution of outer contacts,  introduce a difference in association and disassociation of these contacts in the directions of compression and extension. This difference appears in the evolution equation of the second moment and provides the asymmetry necessary for normal stress differences. Theories of this type produce stress relations that are linear in the strain rate; that is, rate dependent. In contrast, we employ only a short-range repulsive force that is independent of the shear rate. Consequently, our stress relation contains contributions that are independent of rate.

\section{Conclusion}

We have considered a planar extensional flow of a dense layer of spheres in a viscous fluid. In addition to the viscous forces associated with the flow, we assumed that there was a short-range repulsive force between the spheres. We focused on pairs of spheres, assumed that their neighbours translate with the average flow, and required that they be in force and moment equilibrium with each other and their neighbours. We then assumed that the neighbourhoods of pairs with the same orientation were equal to the average over that orientation; this permitted us to write equations for the radial and angular velocity of the relative motion of a single pair as they began and ended an interaction, and the orientation of their line of centres with the axis of greatest compression of the flow. 

 The possible determination of the distribution of near-contacting particles along the trajectory then leads to expressions for the particle shear stress and pressure. Stokesian dynamics simulations provided a value of particle shear stress that permitted the determination of the angle of departure for the trajectories and the separation of near-contacting neighbours. The variation of the shear stress with area fraction suggested the variation of the number of near-contacting 
neighbours per particle with area fraction. With this information, numerical values of the particle distribution and the particle pressure could be calculated. 
 
 The simplicity of the micro-mechanical model makes it possible to understand the influence of the normal and tangential components of the viscous force and the short-range repulsive force on the shape of the trajectory and the distribution of particles along it. It also permits the derivation of a continuum theory for dense suspensions based on the micro-mechanics of equilibrated pairs of particles and a better understanding of those based on phenomenology. However, it is important to note that correlated motions of the particles along the axis of compression in the simulations have no counterpart in the model; these are likely to force approaching particles into trajectories with significantly smaller separations. The small value of $\theta_0$ necessary in the model may be a way for it to incorporate the influence of such particles. Other modelling issues that remain involve the parameters $k$ and ${\bar{s}}$. We believe that it is likely that measurements of the relationship between $k$ and $c$ in three dimensions will be robust and apply to different sharing flows; the determination of ${\bar{s}}$ is less certain, although methods employed by Nazockdast and Morris (2013) may permit its prediction.

The two-dimensional, simple shear flow that was considered earlier by Jenkins and La Ragione (2015) is much more complicated.  In it, there are upstream and downstream trajectories, roughly, on either side of the direction of greatest rate of compression. The exact location of the limiting trajectory is influenced by the rotation of the average flow and, in the calculation, its location is difficult to determine. In the earlier study, the separation,  was taken to be the average separation calculated by Torquato for a dense, equilibrated system of colliding, elastic spheres in two dimensions. This varies between 0.13 and 0.06 as c varies from 0.52 and 0.64. These values are much greater than the value 0.02 employed here. However, in the calculation for simple shear we obtain an excellent agreement between theory and simulation for the shear stress because the length of the particle interaction was taken to be much longer, and probably compensated for the greater separation. Given our experience with the simpler planar extension, we should return to the simple shear flow.
 
The formulation for the two-dimensional planar extensional flow can be easily extended to an axisymmetric extensional flow in three dimensions. The only essential change is in the components of the average strain rate tensor. The three-dimensional calculation can then incorporate particle elasticity and friction to permit a study of shear thickening and comparison with physical experiments.

\section{Appendix A}

Force equilibrium equation for particle $A$ is%
\begin{align}
0 =&6\pi \mu a \te{K}_{\alpha \beta }^{(BA)}v_{\beta }^{(BA)}
-\frac{F_{0}}{s^{(BA)}}\hat{d}_{\alpha}^{(BA)} - 9.54 \pi \mu a^{2}
\left( \hat{t}_{\beta } \te{D}_{\beta \xi } \hat{d}_{\xi} \right) 
\hat{t}_{\alpha}^{(BA)}+\pi \mu a^{2}
\left[ \ln \left( \frac{a}{s^{(BA) }}\right) - 0.96\right] 
\omega^{(A) } \hat{t}_{\alpha }^{(BA) } \notag\\
& + \pi \mu a^{2}\ln \left( \frac{a}{s^{(BA) }}\right) \omega^{(B)}
\hat{t}_{\alpha }^{(BA) }+\sum_{n\neq B}^{N^{(A) }}
\left\{ \frac{3}{\bar{s}}a^{3}\pi \mu 
\left( \te{D}_{\beta \xi} \hat{d}_{\xi }^{(nA)}\hat{d}_{\beta}^{(nA)}\right)
\hat{d}_{\alpha }^{(nA) }+\pi \mu a^{2}\left[
\ln \left( \frac{a}{\bar{s}}\right) -0.96\right] \omega ^{(A) }
\hat{t}_{\alpha }^{(nA)}\right\}  \notag \\
&+\sum_{n\neq B}^{N^{(A)}}\left\{ 2a^{2}\pi \mu \left[ \ln
\left( \frac{a}{\bar{s}}\right) -0.96\right] \left( \te{D}_{\beta \xi }\hat{t}
_{\xi }^{(nA) }\hat{d}_{\beta }^{(nA) }\right) \hat{t}_{\alpha }^{(nA) }
-\frac{F_{0}}{\bar{s}}\hat{d}_{\alpha}^{(nA) }\right\} .  
\end{align}

Similar expression can be written for particle $B$.
The difference between force equilibrium $A$ and $B$ leads to
\begin{align}
0 = & 12\pi \mu a \te{K}_{\alpha \beta }^{(BA) }
v_{\beta }^{(BA)} 
- 2\frac{F_{0}}{s^{(BA)}}\hat{d}_{\alpha }^{(BA)
}- 2\times 9.54\pi \mu a^{2}\left( \hat{t}_{\beta }^{(BA)}
\te{D}_{\beta \xi }\hat{d}_{\xi }^{(BA) }\right) 
\hat{t}_{\alpha}^{(BA) }  \notag \\
&+a^{2}\pi \mu \left[ 2\ln \left( \frac{a}{s^{(BA)}}\right) -0.96\right] S
\hat{t}_{\alpha }^{(BA) }+a^{2}\pi \mu \left[ \ln \left( \frac{a}{\bar{s}}\right)
 -0.96\right] S \varepsilon_{\alpha \beta } Y_{\beta}^{(BA) }  \notag \\
&+2\frac{3}{\bar{s}}a^{2}\pi \mu \te{D}_{\beta \xi }
\te{J}_{\alpha \xi \beta}^{(BA)}+2a^{2}\pi \mu \left[ 
2\ln \left( \frac{a}{\bar{s}}\right) -1.92 \right] 
\te{D}_{\beta \xi }J_{\alpha \xi \beta }^{(BA)}-2\frac{F_{0}}{\bar{s}}
Y_{\alpha}^{(BA)},  \notag
\end{align}
where
\begin{equation*}
S=\omega^{(A)}+\omega^{(B) },
\end{equation*}

\begin{equation*}
\te{J}_{\alpha \xi \beta }^{(BA)}=\sum_{m\neq B}^{N^{(A) }}\hat{d}%
_{\alpha }^{(mA) }\hat{d}_{\beta }^{(mA) }\hat{d}%
_{\xi }^{(mA) },
\end{equation*}
and 
\begin{equation*}
Y_{\alpha }^{(BA) }=\sum_{m\neq B}^{N^{(A) }}\hat{d}%
_{\alpha }^{(mA) }.
\end{equation*}
Moment equilibrium for particle A is 
\begin{align}
0 =&6\pi \mu  a \te{K}_{\alpha \beta }^{(BA)} v_{\beta }^{(BA)} 
\varepsilon_{\alpha \rho }\hat{d}_{\rho }^{(BA)
} - 9.54\pi \mu a^{2}\left( \hat{t}_{\beta }\te{D}_{\beta \xi }
\hat{d}_{\xi}\right) +\pi \mu a^{2}\left[ \ln \left( \frac{a}{s^{(BA) }}\right) -0.96\right] 
\omega^{(A) } \notag\\
&+\pi \mu a^{2}\ln \left( \frac{a}{s^{(BA) }}\right) \omega^{(B)} 
+\varepsilon_{\alpha \rho }\sum_{n\neq B}^{N^{(A)}}
\left\{ \frac{3}{\bar{s}}a^{3}\pi \mu \left( \te{D}_{\beta \xi }\hat{d}_{\xi}^{(nA)}
\hat{d}_{\beta}^{(nA)}\right) \hat{d}_{\alpha}^{(nA)}
+\pi \mu a^{2}\left[ \ln \left( \frac{a}{\bar{s}}\right)
 -0.96 \right] \omega^{(A) }\hat{t}_{\alpha }^{(nA)}\right\} 
\hat{d}_{\rho}^{(nA) }  \notag \\
&+\varepsilon _{\alpha \rho }\sum_{n\neq B}^{N^{(A) }}
\left\{2a^{2}\pi \mu \left[ \ln \left( \frac{a}{\bar{s}}\right) -0.96\right] 
\left(\te{D}_{\beta \xi }\hat{t}_{\xi }^{(nA) }\hat{d}_{\beta }^{(nA)}\right)
\hat{t}_{\alpha }^{(nA)}
-\frac{F_{0}}{\bar{s}}\hat{d}_{\alpha }^{(nA) }\right\} \hat{d}_{\rho }^{(nA) }.  
\end{align}
A similar expression holds for particle $B.$ The sum of moment equilibrium for
particles $A$ and $B$ is
\begin{align}
0 = & 12\varepsilon _{\alpha \beta }\te{K}_{\alpha \mu }^{(BA) }
v_{\mu}^{(BA) } \hat{d}_{\beta }^{(BA) }
-2\times 9.54a\left( \hat{t}_{\mu }^{(BA) } \te{D}_{\mu \xi }\hat{d}_{\xi}^{(BA) }\right) 
 \notag\\
&+a\left[ 2\ln \left( \frac{a}{s^{(BA)}}\right) -0.96\right] S
+a\left[ \ln \left( \frac{a}{\bar{s}}\right) -0.96\right] S\left( k-1\right) \notag \\
&+2a\left[ 2\ln \left( \frac{a}{\bar{s}}\right) -1.92\right] 
\epsilon_{\xi\nu }A_{\nu \mu }^{(BA) }\te{D}_{\mu \xi },  \label{Moment_sum}
\end{align}
where
\begin{equation*}
\te{A}_{\nu \mu }^{(BA) }=\sum_{m\neq B}^{N^{(A) }}\hat{d}_{\mu}^{(mA) }\hat{d}_{\nu}^{(mA)}.
\end{equation*}
In both the difference of force equilibrium and the sum of moment equilibrium, we
have made the following approximations%
\begin{equation*}
\te{A}_{\nu \mu }^{(BA) } = \te{A}_{\nu \mu }^{(AB) },
\end{equation*}
\begin{equation*}
\te{J}_{\alpha \xi \beta }^{(BA)}=-\te{J}_{\alpha \xi \beta }^{(AB)},
\end{equation*}
and
\begin{equation*}
Y_{\alpha }^{(BA) }=-Y_{\alpha }^{(AB) }.
\end{equation*}

The projection of the difference in force equilibrium along the direction
orthogonal to $t_{\alpha }^{(BA) }$ is 
\begin{equation}
3\pi \mu a\frac{a}{s^{(BA) }}\dot{s}^{(BA) }-
2\frac{F_{0}}{s^{(BA) }}+6\pi \mu a^{2}\frac{a}{\bar{s}}
\hat{d}_{\alpha} \te{J}_{\alpha \beta \gamma } \te{D}_{\beta \gamma }
-2\frac{F_{0}}{\bar{s}}Y_{\alpha }
\hat{d}_{\alpha }=0,
\end{equation}
while the component along $t_{\alpha }^{(BA) }$ is
\begin{align}
0 =&12\pi \mu a \te{K}_{\alpha \beta }^{(BA) }
v_{\beta }^{(BA) } t_{\alpha }^{(BA) } - 2\times 9.54\pi \mu a^{2}
\left( \hat{t}_{\beta }^{(BA)} \te{D}_{\beta \xi} 
\hat{d}_{\xi }^{ (BA) }\right)   \notag\\
& + a^{2}\pi \mu \left[ 2\ln \left( \frac{a}{s^{(BA)}}\right) -0.96\right]
S+a^{2}\pi \mu \left[ \ln \left( \frac{a}{\bar{s}}\right) -0.96\right]
S \varepsilon_{\alpha \beta} Y_{\beta }^{(BA)} t_{\alpha}^{(BA) }  \notag \\
&+2\frac{3}{\bar{s}}a^{3}\pi \mu \te{D}_{\beta \xi } \te{J}_{\alpha \xi \beta}^{(BA)}
t_{\alpha}^{(BA)}
+2a^{2}\pi \mu \left[ 2\ln \left( \frac{a}{\bar{s}}\right) - 1.92\right] 
\te{D}_{\beta \xi }\te{J}_{\alpha \xi \beta}^{(BA)}t_{\alpha }^{(BA) }.  
\label{tang_eq_G}
\end{align}%
Using Eq.\,\eqref{tang_eq_G} and Eq.\,\eqref{Moment_sum}, we solve for $S$:

\begin{equation*}
S=-\frac{12b}{\left( 4b-k+1\right) \bar{s}\left[ \ln \left( 1/\bar{s}\right)
-0.96\right] }\sin 2\theta,
\end{equation*}%
or%
\begin{equation}
S=-2c_{2}\sin 2\theta , \label{S_solution}
\end{equation}%
where%
\begin{equation*}
c_{2}=\frac{6b/\left[ \bar{s}\left( 4b-k+1\right) \right] }{\ln \left( 1/\bar{s}\right)-0.96}.
\end{equation*}%

\section{Appendix B}
The Matlab m-files for the numerical solution of the ordinary differential equations and boundary conditions follow.

\includepdf[page={1}]{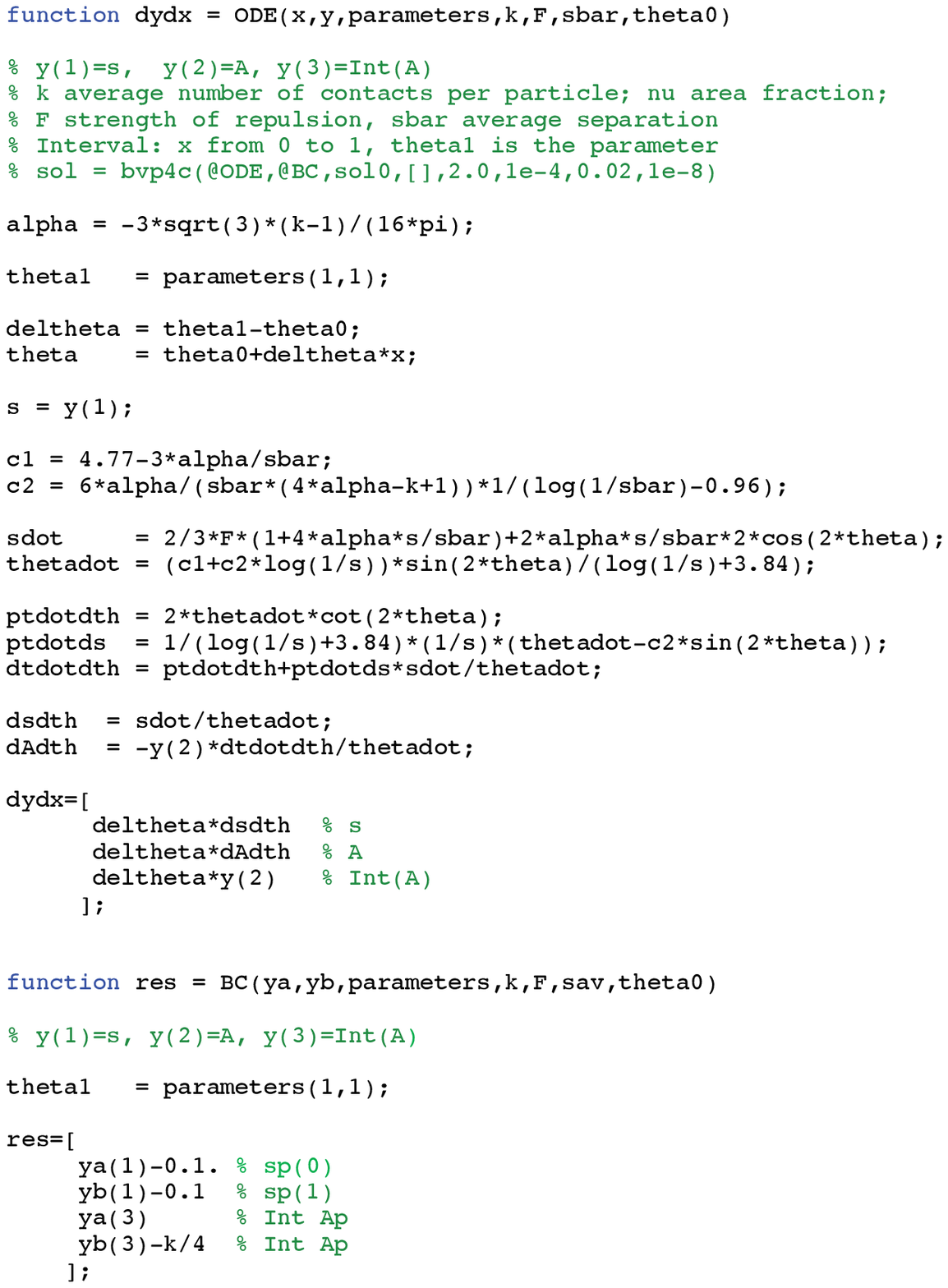}

\vspace{1cm}
This research was partially supported by the National Science Foundation under 
Grant No. NSF PHY-1748958 and by the Gruppo Nazionale della Fisica Matematica (Italy).
R. S. acknowledges the support from the Wenzhou Institute, UCAS.

\section{Declaration of Interests}
 The authors report no conflict of interest.

\end{document}